\documentclass[modern,trackchanges]{aastex63}

\shorttitle{EFR Observed by BBSO/GST}
\shortauthors{Wang et al.}
\usepackage[T1]{fontenc}
\usepackage{textcomp}
\usepackage{colordvi}
\usepackage[normalem]{ulem}
\usepackage{graphicx}
\usepackage{natbib}

\graphicspath{{./figs/}} 
\newcommand{\sm}{$\sim$}
\newcommand{\kms}{$\rm km~s^{-1}$} 

\begin{document}
\title{High-resolution Observations of Small-scale Flux Emergence by GST}
\author{Jiasheng Wang}
\affiliation{Institute for Space Weather Sciences, New Jersey Institute of Technology, University Heights, Newark, NJ 07102-1982, USA; jw438@njit.edu}
\affiliation{Big Bear Solar Observatory, New Jersey Institute of Technology, 40386 North Shore Lane, Big Bear City, CA 92314-9672, USA}
\affiliation{Center for Solar-Terrestrial Research, New Jersey Institute of Technology, University Heights, Newark, NJ 07102-1982, USA}

\author{Chang Liu}
\affiliation{Institute for Space Weather Sciences, New Jersey Institute of Technology, University Heights, Newark, NJ 07102-1982, USA; jw438@njit.edu}
\affiliation{Big Bear Solar Observatory, New Jersey Institute of Technology, 40386 North Shore Lane, Big Bear City, CA 92314-9672, USA}
\affiliation{Center for Solar-Terrestrial Research, New Jersey Institute of Technology, University Heights, Newark, NJ 07102-1982, USA}

\author{Wenda Cao}
\affiliation{Institute for Space Weather Sciences, New Jersey Institute of Technology, University Heights, Newark, NJ 07102-1982, USA; jw438@njit.edu}
\affiliation{Big Bear Solar Observatory, New Jersey Institute of Technology, 40386 North Shore Lane, Big Bear City, CA 92314-9672, USA}
\affiliation{Center for Solar-Terrestrial Research, New Jersey Institute of Technology, University Heights, Newark, NJ 07102-1982, USA}

\author{Haimin Wang}
\affiliation{Institute for Space Weather Sciences, New Jersey Institute of Technology, University Heights, Newark, NJ 07102-1982, USA; jw438@njit.edu}
\affiliation{Big Bear Solar Observatory, New Jersey Institute of Technology, 40386 North Shore Lane, Big Bear City, CA 92314-9672, USA}
\affiliation{Center for Solar-Terrestrial Research, New Jersey Institute of Technology, University Heights, Newark, NJ 07102-1982, USA}

\begin{abstract}    
    Recent observations demonstrated that emerging flux regions, which constitute the early stage of solar active regions, consist of emergence of numerous small-scale magnetic elements. They in turn interact, merge, and form mature sunspots. 
    However, observations of fine magnetic structures on photosphere with sub-arcsecond resolution are very rare due to limitations of observing facilities. 
    In this work, taking advantage of the high resolution of the 1.6~m Goode Solar Telescope, we jointly analyze vector magnetic fields, continuum images, and H$\alpha$ observations of NOAA AR 12665 on 2017 July 13, with the goal of understanding the signatures of small-scale flux emergence, as well as their atmospheric responses as they emerge through multiple heights in photosphere and chromosphere. 
    Under such a high resolution of 0$\farcs$1 to 0$\farcs$2, our results confirm two kinds of small-scale flux emergence: magnetic flux sheet emergence associated with the newly forming granules, and the traditional magnetic flux loop emergence. 
    With direct imaging in the broadband TiO, we observe that both types of flux emergence are associated with darkening of granular boundaries, while only flux sheets elongate granules along the direction of emerging magnetic fields and expand laterally. 
    With a life span of 10\sm15~minutes, the total emerged vertical flux is in order of ~$10^{18}$~Mx for both types of emergence. 
    The magnitudes of the vertical and horizontal fields are comparable in the flux sheets, while the former is stronger in flux loops. 
    H$\alpha$ observations reveal transient brightenings in the wings in the events of magnetic loop emergence, which are most probably the signatures of Ellerman bombs. 
\end{abstract}


\section{introduction}
    Flux emergence, through which magnetic fields are transported to the solar atmosphere from  sub-surface, is considered to be generated by convective motions and aided by magnetic buoyancy  in solar interior \citep{1968IAUS...35...95S,1974IAUS...56...35S}.
    Flux emergence in different scales is important for energy release in different forms, inlcuding small-scale brightenings and large-scale solar eruptions.
    The typical  scenario of emergence is that magnetic fields are twisted underneath the  photosphere due to flows and rise to form an $\Omega$-loop due to magnetic buoyancy \citep{1977ARA&A..15...45P,2001ApJ...554L.111F}.
    Observations of emissions in the solar atmosphere such as in UV/EUV provide evidence that energy may be released due to reconnection during the process of emergence. 
    The magnetic fields emerging through convection zone are not constrained to rise in an aligned orientation with the pre-existing field, so the magnetic reconnection is expected to occur between the emerging fluxes and pre-existing fluxes. 
    Overall, on the large scale in the solar photosphere, the orientation of emerging fields is roughly aligned with the direction connecting paired polarity spots \citep{2011PASJ...63.1047O,2012ApJ...759...72C}.
    
    Taking advantage of high-resolution (\sm0$\farcs$3) observations, \cite{2002ApJ...569..474D} found that magnetic concentrations emerge within the granule interior and quickly (\sm10--15~minutes) disperse following granule flows. The author speculates such flux emergence initiates with horizontal magnetic structures. 
    The study of \cite{2007A&A...467..703C} supported these findings from the simulation perspective.
    They found that emerging magnetic elements with sufficiently high field strength can also impact on the granular structure. 
    The elongated granule and intergranular lane darkening are reported on the photosphere with observations in visible wavelengths \citep{1985SoPh..100..397Z, 2011ApJ...740...82L, Yang_2016}. 
    On the other hand, the recent study of \cite{2019A&A...622A.168C} showed that while small magnetic elements are advected to upper layers on the surface through normal convection, emergent magnetic fields with B$\geqslant$50~G tend to in turn induce the photospheric motions by governing the plasma flows. The authors also found that such emergent-flux-related flow fields change velocity distributions as well as granule elongation.

    Besides the dynamic magnetic characteristics observed on the photosphere, variations of brightness from continuum images provide clear  indications of magnetic flux emergence.
    \cite{2012arXiv1207.6418Y} found that small-scale flux emergences have associated bright points on the photosphere, mostly inside solar granulation, in which the field emerges at a size scale less than 1--2~Mm \citep[e.g][]{1996ApJ...460.1019L,2002ApJ...569..474D}. They suggested that the emergence of relatively strong fields create bright points at the footpoints of magnetic loops, which intrude into intergranular lanes.
    Ellerman bombs (EBs) \citep{1917ApJ....46..298E}, the bright signatures essentially observed in H$\alpha$ wings, are found at locations where magnetic elements with opposite polarities are close to each other. They are likely linked with the dips of the serpentine magnetic field through the surface \citep{2004ApJ...614.1099P,2013A&A...557A.102B}.
    The previous studies of EBs conclude that such photospheric heating processes are caused by photospheric reconnection of strong opposite-polarity field and are not directly associated with chromosphere and transition region dynamics \citep[e.g][]{Watanabe_2011,Vissers_2013,2015ApJ...812...11V}.

    Since the first observational report of granulation scale emergence events \citep{2002ApJ...569..474D}, high-resolution polarimetric observations focus more frequently on small-scale flux emergence events together with observations of flow motions.
    In the high-resolution (\sm0$\farcs$32) observation of small-scale flux emergence reported by \cite{2007ApJ...666L.137C}, by using analysis of local thermodynamic equilibrium (LTE) inversion of full Stokes measurements, the author found horizontal field emergence prior to the appearance of vertical flux elements in the typical granulation time scales (10 minutes). 
    With the advance of observational technology, the existence of flux loops have been witnessed \citep[e.g.,][]{2009ApJ...700.1391M,2010ApJ...720..454T}.
    By implementing magnetohydrodynamic (MHD) simulations of magnetoconvection, \cite{2018ApJ...859L..26M} detected two types of flux emergence events: magnetic loop emergence and flux sheet emergence.  
    In previous observations of the flux loop emergence with Hinode \citep[e.g.,][]{2007ApJ...666L.137C,2009ApJ...700.1391M,2017ApJS..229...17S}, the authors summarized physical characteristics of the emergence: the horizontal field enhances within a well-established granule structure followed by emerged vertical fields drifting in intergranular lanes. The vertical field elements are connected by horizontal magnetic patches. 
    Recent studies by \cite{2017ApJS..229....3C} and \cite{2019A&A...622L..12F} have reported the flux-sheet emergence events, which have different signatures from flux loop emergence. Instead of evolving within granules, the horizontal field enhances together with the expansion of a granule. This forms an organized sheet-like mantle that spans both in the emerging direction and to sides. 
    The sheet covers the entire granule, and the emerged longitudinal flux in footpoints is also in the order of $10^{18}$~Mx. 

    In this paper, we study the magnetic field structure and evolution during the flux emergence in the NOAA active region (AR) 12665 on July 13, 2017.
    Taking advantage of the exceptionally high resolution of the 1.6~m off-axis Goode Solar Telescope \citep[GST;][]{2012SPIE.8444E..03G} at Big Bear Solar Observatory (BBSO), we are able to observe fine magnetic structures on sub-arcsecond level (0$\farcs$1 to 0$\farcs$2) and study the magnetic properties in both flux emergence scenarios as described above.
    We also investigate photospheric and chromospheric brightness variation, especially Ellerman bombs, associated with the small-scale flux emergence. 
    The structure of this paper is as follows: We introduce our observations and data processing methods in Section \ref{data}; In Section \ref{results}, we present analyses of observational results; Key findings are summarized and discussions are presented in Section \ref{sum}.

\section{Observations and data processing}\label{data}
    As the Sun enters the activity minimum, observations of ARs are less often obtained. 
    On the other hand, with the routine operation of GST at BBSO, the quiet Sun and less complicated ARs are more feasible targets. 
    Aided by the high-order adaptive optics system with 308 sub-apertures \citep{2014SPIE.9148E..35S} and completion of the second generation of spectro-polarimetric instrument -- the Near Infra-Red Imaging Spectro-polarimeter (NIRIS) \citep{2012ASPC..463..291C}, BBSO/GST obtained observations near the main magnetic polarity inversion line (PIL) of NOAA AR 12665 (31$^{\circ}$W, 6$^{\circ}$S) during \sm20:16--22:42 UT on 2017 July 13. 
    Under the excellent seeing conditions, the observations achieved diffraction-limited imaging with a resolution of 0$\farcs$1 to 0$\farcs$2.
    The data includes spectro-polarimetric observations of full sets of Stokes measurement at the Fe I 1564.8~nm line (0.25~\AA\ bandpass) by NIRIS with a round field of view (FOV) of 80\arcsec\ at 0$\farcs$24 resolution and 56~s cadence, 
    Fabry-P\'erot spectroscopic observations around H$\alpha$ line center at $\pm$ 1.0, $\pm$ 0.6, $\pm$ 0.4, and 0.0~\AA\ (0.08~\AA\ bandpass) by the Visible Imaging Spectrometer (VIS) with a 70\arcsec\ circular FOV at 0$\farcs$1 resolution and 33~s cadence, and images in TiO (705.7~nm; 10~\AA\ bandpass) by the Broad-band Filter Imager with a 70\arcsec\ circular FOV at 0$\farcs$1 resolution and 15~s cadence.
    TiO and H$\alpha$ observations achieved a diffraction-limited resolution in the order of 0$\farcs$1 with speckle-masking image reconstruction  \citep{2008A&A...488..375W}, while NIRIS achieves a spatial resolution of 0$\farcs$24 without speckle reconstruction.
    
    Alignment among H$\alpha$ images, TiO images, and magnetograms are processed by matching the most stable sunspot and plage features in the FOV. 
    After data noise deduction, the essential vector magnetograms from NIRIS are obtained through Stokes inversion based on Milne-Eddington approximation \citep[see Methods in][]{2017NatAs...1E..85W} and aligned by using interpolation to achieve sub-pixel precision.
    Vector magnetograms in the local coordinates were deduced after removing the 180$^{\circ}$ azimuthal ambiguity with the AUTO-AMBIG code by \cite{2009ASPC..415..365L,2009SoPh..260...83L}, which is an optimized dis-ambiguation method originally intended for $Hinode$ vector data.
    It uses the minimum energy algorithm \citep{1994SoPh..155..235M} to find a minimum of field divergence ($\nabla\cdot$B) and current density (J) in the FOV. 
    To assist in tracking magnetic elements and quantification of magnetic flux, we applied the Southwest Automatic Magnetic Identification Suite (SWAMIS) \citep{2007ApJ...666..576D}, which is a demonstrated technique for magnetic identification and tracking. 
    Here we set the threshold of the vertical magnetic field to 100~G.  Based on visual inspection, this threshold allows us to include as many detected magnetic elements as possible while maintaining a high S/N ratio.

\section{Results and Analysis}\label{results}
\begin{figure}
    \epsscale{1.2}
    \plotone{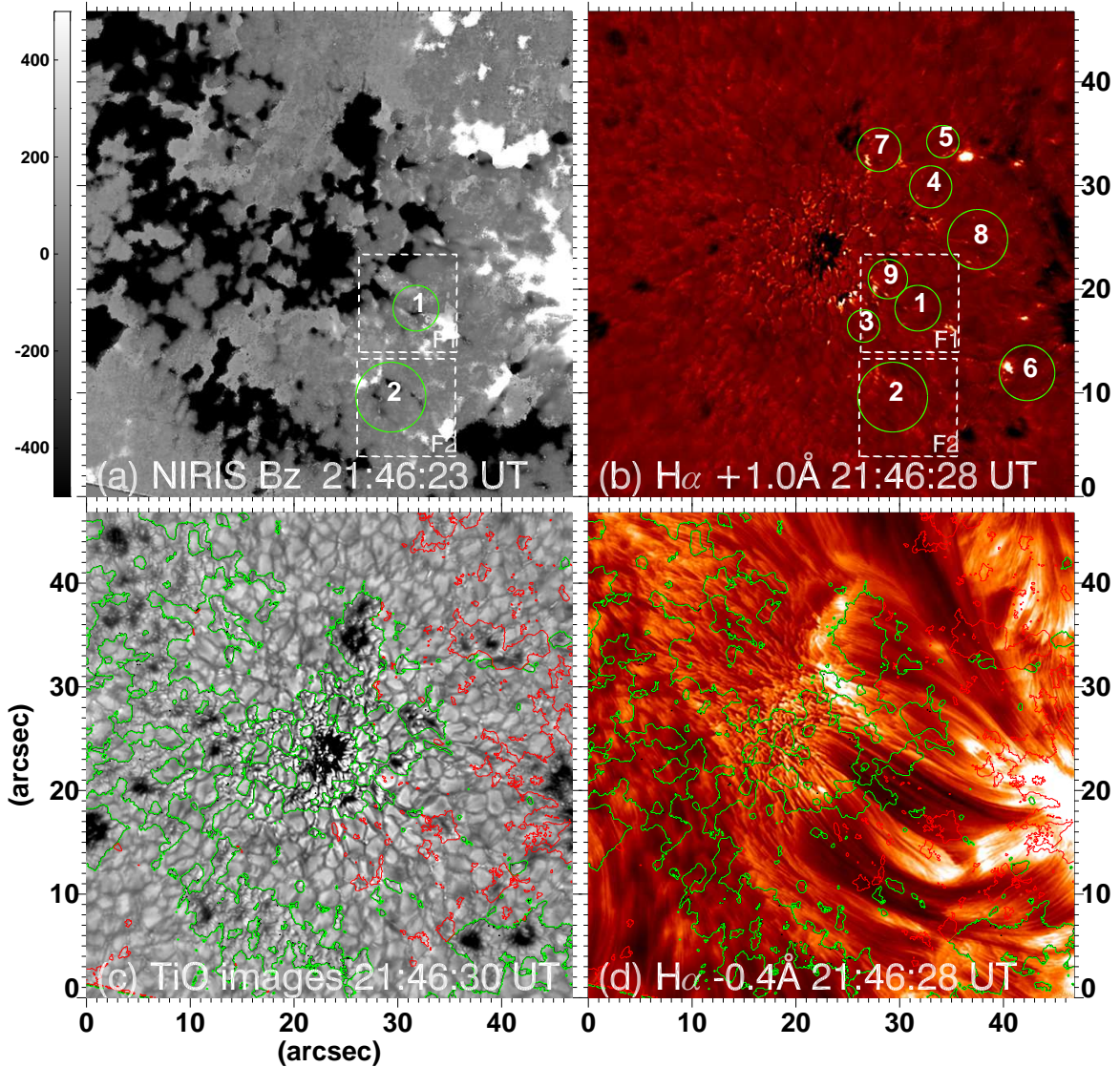}
    \caption{\textbf{Overview of the emergence observations.} Multi-wavelength observations from GST at 21:46 UT is displayed in the figure. 
    Panel (a) shows vertical magnetic field map, whose magnitude is represented in gray scale with black (white) meaning negative (positive) polarity. Grayscale of the vertical field map saturates at $\pm$500~G.
    Panel (b) and (d) show H$\alpha$ images at +1.0 and $-$0.4~\AA, respectively. Green circles indicate regions of observed emergence events and white dashed boxes (F1 and F2) indicate FOV of \ref{fig:f2} and \ref{fig:f4}
    Panel (c) is TiO image that shows photospheric structures.
    Panel (c) and (d) are overplotted with vertical field contours of $\pm$150~G, in which green (red) indicate negative (positive) values.An animation of the GST multi-wavelength observations is available in the online Journal.
    The animated Figure, which includes the same panels (a)-(d) shown here, runs from 20:27 to 22:35~UT. 
    The animated images are not annotated except for the F1 and F2 FOV. \label{fig:f1}}
\end{figure}

    GST observation was centered at the flare productive NOAA AR 12665 at (432\arcsec,-164\arcsec). The AR is classified as the $\beta\gamma$ magnetic configuration. 
    Figure \ref{fig:f1} and the online animations show an overview of the AR in magnetograms, TiO images, and H$\alpha$ images at +1.0 and $-$0.4~\AA. 
    During the period of observation, there is obvious magnetic flux emergence of opposite polarities at the main PIL. 
    Emerging magnetic elements actively diverge from the PILs and eventually merge into the nearby sunspots. 
    The TiO visible images clearly show that granules near the PIL exhibit elongating patterns. Such evolving granular structures are typical photospheric signatures of flux emergence. 
    Simultaneous magnetic field measurements taken by NIRIS reveal an enhanced horizontal field accompanied by the elongating granules. 
    Concentrated magnetic elements of opposite polarities are located at the two ends of the central region with the enhanced horizontal field.
    
    In Figure \ref{fig:f1}(b), the H$\alpha$ image clearly exhibits brightenings at the footpoints of the emerging fibrils associated with the new flux emergence and growing pores. 
    The green circles outline the locations of small-scale flux emergences labeled 1 to 9. The diameters of circles correspond to the size of the associated granules in TiO images. 
    The white dashed boxes F1 and F2 indicate the regions of events that we will discuss in Sections \ref{sub.1} and \ref{sub.2}.
    The vertical component of magnetic fields is shown in Figure \ref{fig:f1}(a), which saturates at $\pm$500~G. 
    From the online TiO movie, one can see that the magnetic flux is transported to the photosphere through individual episodes in the scale of granules during flux emergence. Subsequently, the Sun's pore areas are expanded as the same polarity fluxes are merged to them.  
    From H$\alpha$ off-band images, flows in dark fibrils are observed streaming toward or away from the concentrated magnetic footpoints. 
    
    During the observation time window, we identified eight good events (see Table \ref{tab:list}) of small-scale flux emergence that have high-quality data in all wavelengths obtained. The magnetic topology of event 5 can not be clearly interpreted because the magnetograms lack the accuracy of azimuthal disambiguation in this event area. For a similar reason, we exclude some emergence events seen in continuum images. 
    Each of them has an emerged total unsigned flux in the order of $10^{18}$~Mx and shows prominent  magnetic structure changes on the photosphere. 
    The observed lifetime of these emergence events is \sm10~minutes, which is on the same scale as the lifetime of granulation. Thus the observed flux emergence events are considered as granular-sized magnetic flux emergence. Different magnetic characteristics are observed in these small-scale flux emergence events with high-resolution data.
    In the case studies of observed emergent events, we are able to distinguish two different types of flux emergence processes, i.e., flux sheet emergence and flux loop emergence \citep[e.g][]{2009ApJ...700.1391M,2017ApJS..229....3C,2019A&A...622L..12F}. 
    In the case studies of the observed emergent events, the two types of flux emergence events are categorized based on geometric properties of the field evolution and correspondent structure changes.

\subsection{Detailed Study of a Flux Sheet Emergence}\label{sub.1}

    Since the observed emergence events are visible in granule-sized scale and often adjacent to actively evolving granules, the clear event episodes are selected manually after implementing the SWAMIS feature tracking method. 
    In the five identified events of flux sheet emergence among all eight selected events, an enhanced horizontal field is seen to emerge within small granules as well as in the intergranular dark lane that later forms a newly emerged granule cell. 
    The emerging horizontal field expands its boundaries in the directions both along and across the field lines while the field lines within granule cells are aligned between concentrated footpoints of opposite polarities. 
    We also found that on average the horizontal magnetic field strength (265~G) is comparable with the vertical field (272~G) in the emergent area as both are enhanced during sheet emergence. Despite small variations in individual cases, the emerging flux expands its front at a speed of 1.5~\kms ($\pm$0.55~\kms). In event 1 we observed the highest speed of emerged footpoints at 2.1~\kms, and in event 8 we observed the lowest speed at 0.8~\kms.  
    TiO images show that the photospheric granular structures associated with emerged footpoints' separations undergo expansion during the flux emergence process, then follow the typical life cycle of photospheric granulation. 
 
\begin{figure}
    \epsscale{1.2}
    \plotone{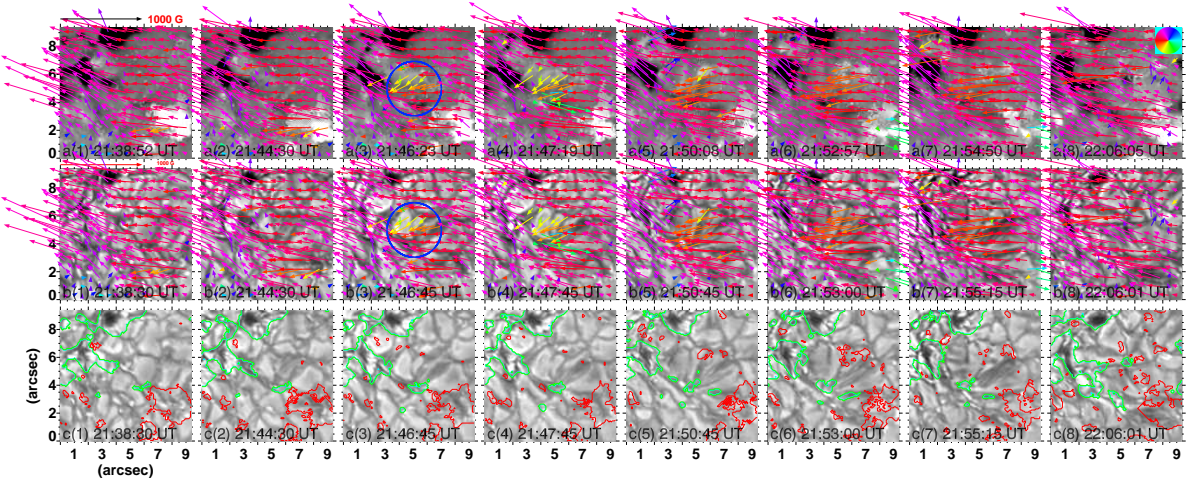}
    \caption{\textbf{Temporal evolution of emergence event 1.}
    The figure shows snapshots of emergence event 1 from 21:46~UT to 22:06~UT. 
    Panels (a) show vertical field superimposed with horizontal field vectors, whose directions are represented by vector colors and magnitude is represented by length.
    Panels (b) show horizontal field on top of TiO images.
    Panels (c) show TiO images overlied with vertical magnetic elements, the red (green) contours represent positive (negative) magnetic elements at level of 150 G.
    The blue circle in Figures \ref{fig:f2}(a3) and (b3) indicates the location of emergent flux sheet with correspondent expanding granule in the background. An animation of emergence event 1 is available in the online Journal.
    The animated Figure, which includes the same panels (a)-(c) shown here, runs from 21:32 to 22:16~UT. \label{fig:f2}}
\end{figure}

    By reviewing the time-lapse movies of event 1 in multi-wavelengths, we identified continuous flux emergence and evolving granulation structure, which belong to the flux-sheet emergence type. 
    The event 1 lasts \sm50 minutes, during which the TiO images and horizontal magnetic field maps clearly show two stages of the emergence process. 
    Figure \ref{fig:f2} shows the temporal evolution of magnetic and continuum structures  of this event. 
    Figures \ref{fig:f2}(a1--a8) present snapshots of image sequence from 21:46~UT to 22:06~UT of vertical field superimposed with horizontal field vectors, whose directions are represented by colors and magnitude is represented by arrow length. The cutoff value of the horizontal field vectors is 100~G. 
    Figures \ref{fig:f2}(b1--b8) show TiO images overlaid with the same horizontal field vectors as in Figures \ref{fig:f2}(a1--a8). 
    From Figures \ref{fig:f2}(b3--b4), we clearly observe that the disoriented field vectors overlap entirely an expanding granule.
    Figures \ref{fig:f2}(c1--c8) present TiO images superimposed with vertical magnetic elements, with the green (red) contours representing negative (positive) magnetic field at a magnitude of 150~G. 
    The concentrated magnetic elements are seen to be located at the intergranular boundaries as new fluxes emerge to the photosphere \citep{Jin_2008}. 
    In the region where flux emergence occurs (blue circle in Figures \ref{fig:f2}(a3) and (b3)), concentrated magnetic elements divert along the intergranular lanes near the western edge of the region and eventually merge with pores of the same polarities (as shown throughout Figures \ref{fig:f2}(a1--a8)). 
    For a very short period of \sm10~minutes (as seen in first four columns in Figure \ref{fig:f2}), a granule cell appears near the edge (centered at [X,Y]\sm[5\arcsec,5\arcsec]) of a pre-existing granule and grows in the circled region with the overlying horizontal field emerging in the direction nearly perpendicular to the predominant direction of ambient fields.
    The translational motion of negative magnetic elements along the intergranular lane is observed at the western side of the circled area in Figures \ref{fig:f2}(a5--a7) and (c5--c7).
  
\begin{figure}
    \epsscale{1.2}
    \plotone{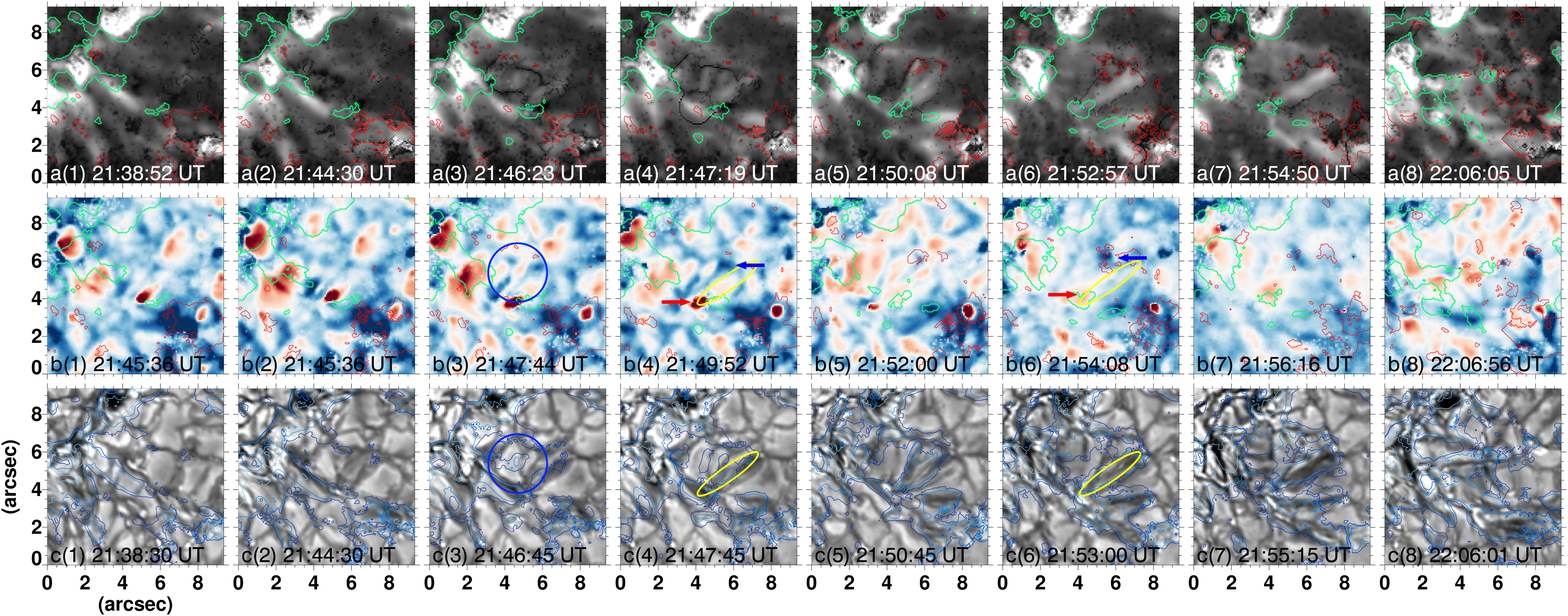}
    \caption{\textbf{Horizontal field and dopplermaps in event 1.} 
    Panels (a) show horizontal field map superimposed with vertical field contours at level of 150 G. 
    Panels (b) show upflows (downflows) of Dopplergrams in blue (red) color. The line-of-sight component the correspondent velocity is in range of $\pm$3.0~\kms. 
    Panels (c) present TiO images superimposed with horizontal field contours at levels of 200~G and 400~G, indicated by dark and light blue, respectively. 
    The green (red) contours in (a) and (b) represent magnetic elements of negative (positive) polarity at level of 150 G.
    Blue circle in (b3) and (c3) indicate the location of expanding granule. intergranular lane is outlined with ellipse in (c4) and (c6). 
    Blue (red) arrows in (b4) and (b6) indicate strong Doppler blue-shift (red-shift) at footpoints. An animation of the horizontal field and dopplermaps in event 1 is available in the online Journal.
    The animated Figure, which includes the same panels (a)-(c) shown here, runs from 21:32 to 22:16~UT. \label{fig:f6}}
\end{figure}

    The background field in the studied region is approximately in the east-west direction. 
    At the start of the time sequence in Figure \ref{fig:f2}, granulation is accompanied with the growth of a new granule cell. Along with the disoriented granule expansion occurrence (Figure \ref{fig:f2}(b3)), the accompanying horizontal field emerges in an organized direction different from the pre-existing field. The newly emerged horizontal field extends its boundary as it enhances in 8 minutes. 
    In Figure \ref{fig:f6}, enhanced horizontal field patches are observed at \sm21:47~UT and two minutes later, the enhanced fields reach the boundary of the co-spatial granule, where vertical magnetic fields concentrate into footpoints as indicated by the red and green contours (shown in Figures \ref{fig:f6}(a4--a6)). 
    The noticeable enhancement of the horizontal field at the granule's west edge as seen in Figure \ref{fig:f6}(a5) is associated with a developing dark lane.
    When the vertical field is concentrated to the extended intergranular lanes as shown in Figure \ref{fig:f2}(a6), the horizontal field continues to enhance (Figures \ref{fig:f6}(a5--a7)). 
    The most prominent enhancement covers the elongated granule and intergranular dark lane. From the dopplergrams in Figures \ref{fig:f6}(b1--b8), both upflows and downflows are observed in the flux sheet area (centered at \sm[4\arcsec,5\arcsec] in Figures \ref{fig:f6}(b3--b4)).
    Strong Doppler blue-shifts (red-shifts) with upflow (downflow) velocity up to 1.8~\kms\ are observed at the positive (negative) footpoints in the intergranular lanes (centered at \sm[6\arcsec,6\arcsec] in Figure \ref{fig:f6}(b6)). Very weak blue-shifts are seen within the granular cell (centered \sm[4\arcsec,6\arcsec] in Figure \ref{fig:f6}(b6)), where the average Doppler upflow velocity is \sm0.4~\kms. 
    This is roughly two times smaller than that of emerging flux in the previous study of \cite{2017ApJS..229....3C}, and is also smaller than the average upflows (downflows) of 0.64 (0.49)~\kms\ as found by \cite{2017ApJ...849....7O}. 
    
\begin{figure}
    \epsscale{1.2}
    \plotone{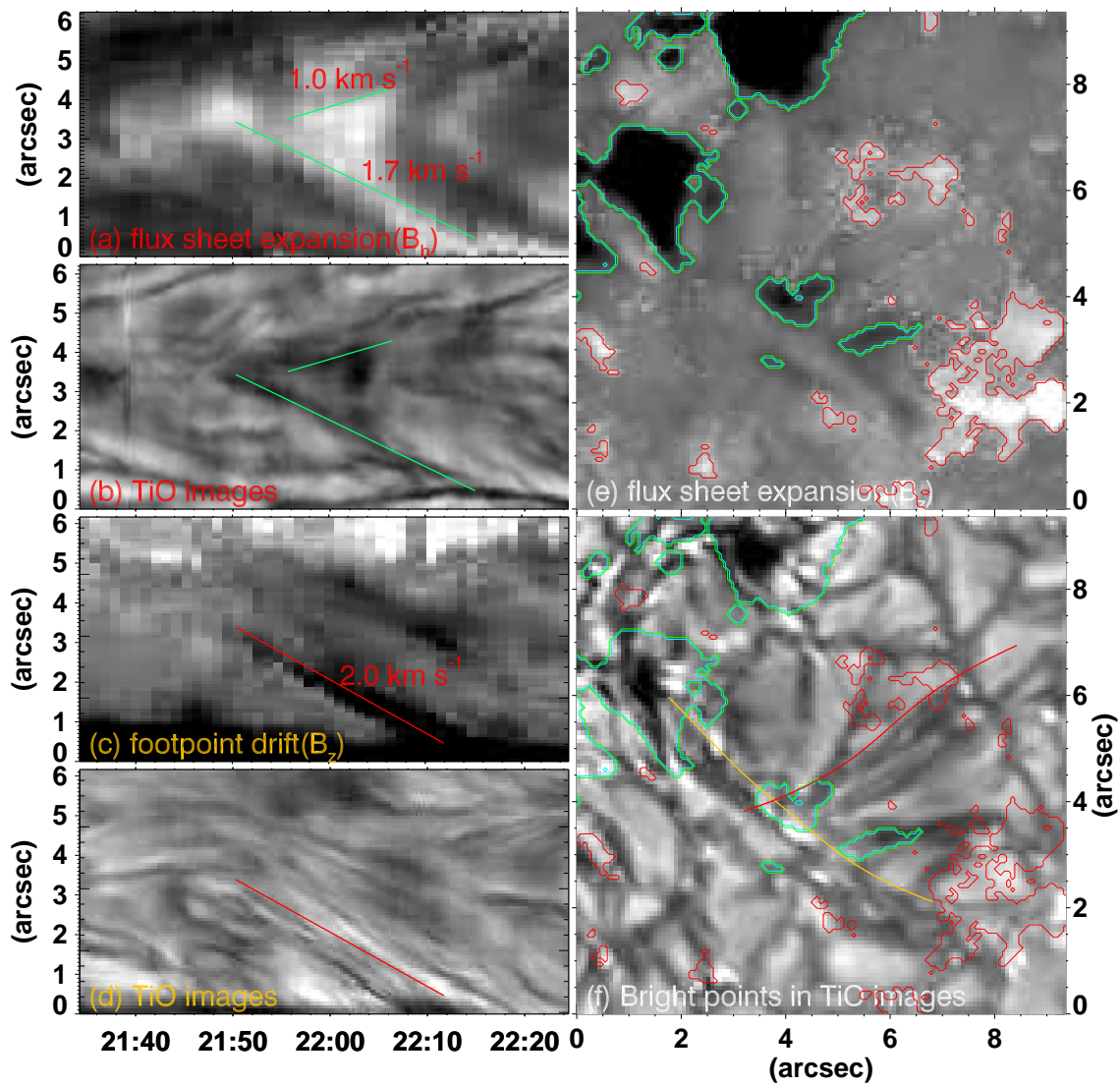}
    \caption{\textbf{Time-space diagram of event 1.} 
    Panels (a) and (b) show time-space diagrams of horizontal field and TiO along the red slit as shown in (f), which correspond to flux sheet emergence stage. Green lines in (a) and (b) trace the expanding granule. 
    Panels (c) and (d) show time-space diagrams of vertical field and TiO along the yellow slit as shown in (f), which represent negative footpoint motions in the intergranular lane. Red lines in (c) and (d) trace and are used to estimate speed of motion of the magnetic element.
    Green (red) contours in (e) and (f) outline the concentrated negative (positive) magnetic elements. \label{fig:f3}}
\end{figure}

    To further analyze the magnetic evolution associated with flux emergence, we present the time-distance diagrams of horizontal field and TiO image in Figures \ref{fig:f3}(a)--(d), which display the time-distance evolution of two slits across the flux sheet and along negative footpoint trail indicated in Figure \ref{fig:f3}(f) as red and yellow curves, respectively.
    Figure \ref{fig:f3}(a) clearly shows the enhancement of horizontal field in the expanding granule, in which the separating bright lanes represent the emerging horizontal field with a magnitude over 150~G.
    The associated bi-directional extending granule boundaries are presented in Figure \ref{fig:f3}(b) based on TiO observations. 
    The observations show that the emergence in the granulation starts at 21:46~UT, when the horizontal field starts to increase from the background field and fills the granule interior. 
    The ongoing emergence lasts \sm15 minutes before dark intergranular lanes form in place at \sm22:02~UT. 
    The concentrated footpoints (as indicated by the green contours in Figure \ref{fig:f3}(e)) at the boundary continue to evolve with an expansion speed of \sm1.7~\kms. 
    Associated with the horizontal field emergence in the transverse direction, the front of the growing granule as indicated by TiO dark lanes (seen in Figure \ref{fig:f3}(b)) expands at the same speed.  
    The time-distance diagram (shown in Figure \ref{fig:f3}(c)) along the yellow slit indicates that the motion of the negative magnetic element resides in the intergranular lane. Its speed of motion along the slit is 2~\kms. 
    Figure \ref{fig:f3}(d) shows the co-spatial TiO evolution in the intergranular lane. 
    Although granular boundaries are observed as dark lanes in TiO images, we find that the concentrated magnetic elements are associated with transient TiO bright points. The negative magnetic elements and the co-spatial TiO bright points drift together along the intergranular lane.
    The horizontal field in the flux sheet emergence event 1 increases throughout the 20 minutes evolution, reaching up to 450~G. The newly emerged vertical flux at the negative footpoint is 1.3$\times$~$10^{18}$~Mx.

\subsection{Detailed Study of a Flux Loop Emergence}\label{sub.2}
    On the other hand, in regions where events of emerging granules take place less often, we observed dumbbell-like features in magnetograms representing flux loop emergence events, with two ends of loops rooted in opposite magnetic polarities. 
    The emergence of magnetic concentrations originates in the boundaries of neighboring granules and then the emerged elements move along the magnetic network. A relatively weak field connects the two emerged footpoints. It is seen that the emerged magnetic footpoints do not alter the overall evolution of their nearby granules. As shown in the online movies, the passage of flux loop footpoint motions shifts following the nearby granule emergence and decay, which means that the merged flux loop does not dominate the local magnetic field and structure evolutions.
    {By comparing averaged field strength we found that the emerged vertical field is 326~G, which is \sm120~G (60\%) higher than the emerged horizontal field (\sm200~G).}
    We observe TiO and H$\alpha$ brightenings more often in this flux loop type of emergence. In particular, all three events are seen to be spatially associated with H$\alpha$ brightenings near the emerged magnetic footpoints.

    Event 2 (indicated by the box F2 in Figure \ref{fig:f1}) is one of the distinctive magnetic loop type of flux emergence in our observations, in which the emerging magnetic footpoints travel in the network along intergranular dark lanes and are connected by an arched magnetic field. 
    With the aid of H$\alpha$ off-band images, we also observe Ellerman bombs at the negative polarity  footpoint and additional brightenings at the central location in this event.  
    
\begin{figure}
    \epsscale{1.2}
    \plotone{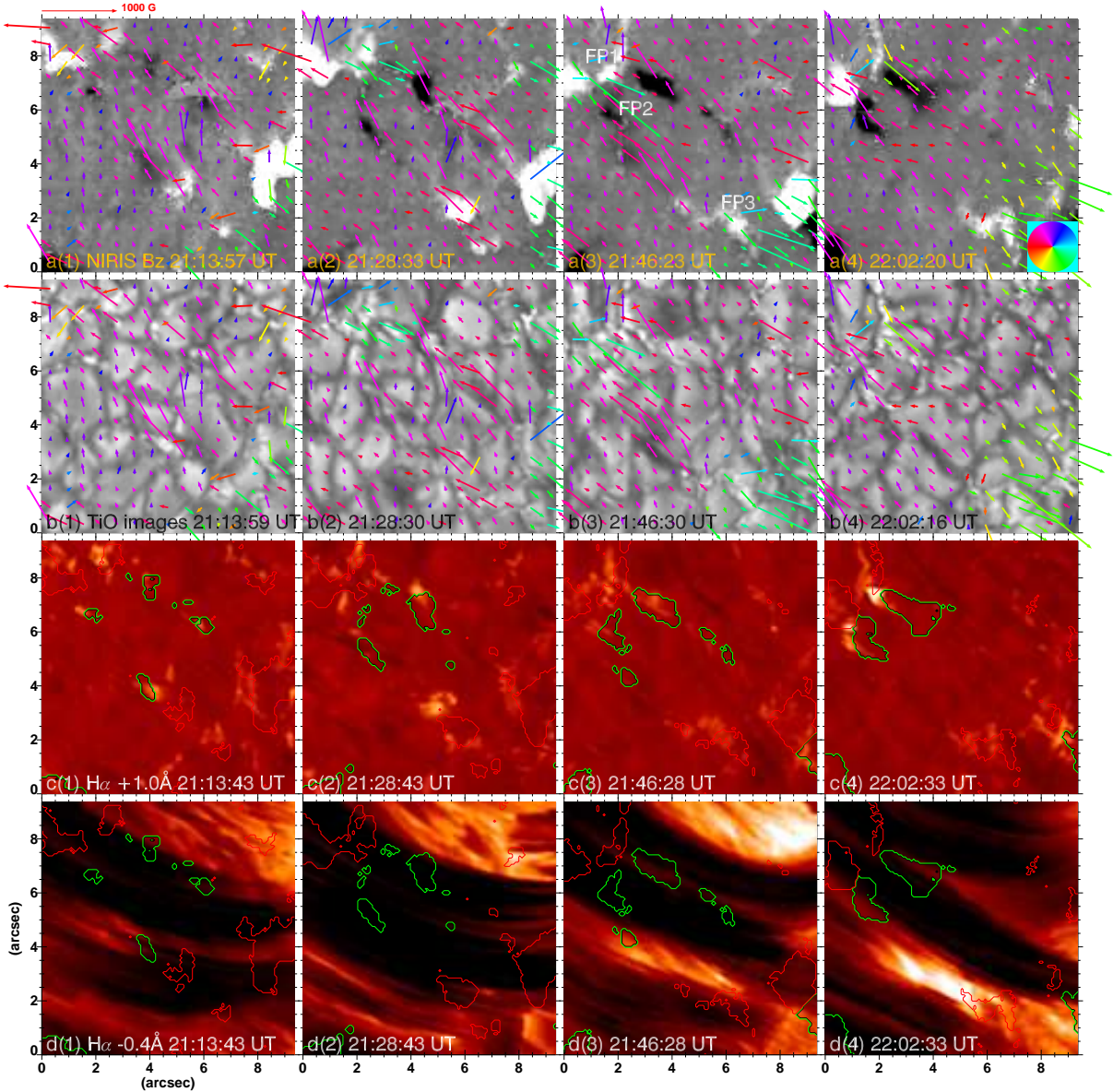}
    \caption{\textbf{Temporal evolution of emergence event 2.} 
    The figure shows snapshots of emergence event 2 from 21:13~UT to 22:02~UT.
    Panels (a) show vertical field superimposed with horizontal field vectors, whose directions are represented by vector directions and magnitude is represented by length.
    Panels (b) show TiO images overlied with horizontal field vectors.
    Panels (c) and (d) show H$\alpha$ images at +1.0 and $-$0.4~\AA, the green (red) contours represent negative (positive) magnetic elements at level of 150 G. An animation of emergence event 2 is available in the online Journal.
    The animated Figure, which includes the same panels (a)-(d) shown here, runs from 21:03 to 22:16~UT. \label{fig:f4}}
\end{figure}

    Figure \ref{fig:f4} shows the temporal evolution of the elementary flux emergence that forms a magnetic loop configuration using the magnetic and continuum observations. 
    In the snapshots of vector magnetic field maps (as shown in Figures \ref{fig:f4}(a1--a4)), horizontal field vectors are superimposed on vertical fields and are also overplotted on TiO images (Figures \ref{fig:f4}(b1--b4)). 
    The direction of the horizontal field is indicated by the direction of the arrow and displayed in different colors for each direction, and the positive (negative) vertical field is indicated by the white (black) background. 
    Figures \ref{fig:f4}(c1--c4) and Figures \ref{fig:f4}(d1--d4) show H$\alpha$ images at +1.0 and $-$0.4~\AA, respectively, with the overplotted green (yellow) contours representing the negative (positive) magnetic elements at the level of 150~G.
    From the image sequence and the online movie, we can see consecutive episodes of flux emergence during the time of observation in the event 2 region. 

\begin{figure}
    \epsscale{1.2}
    \plotone{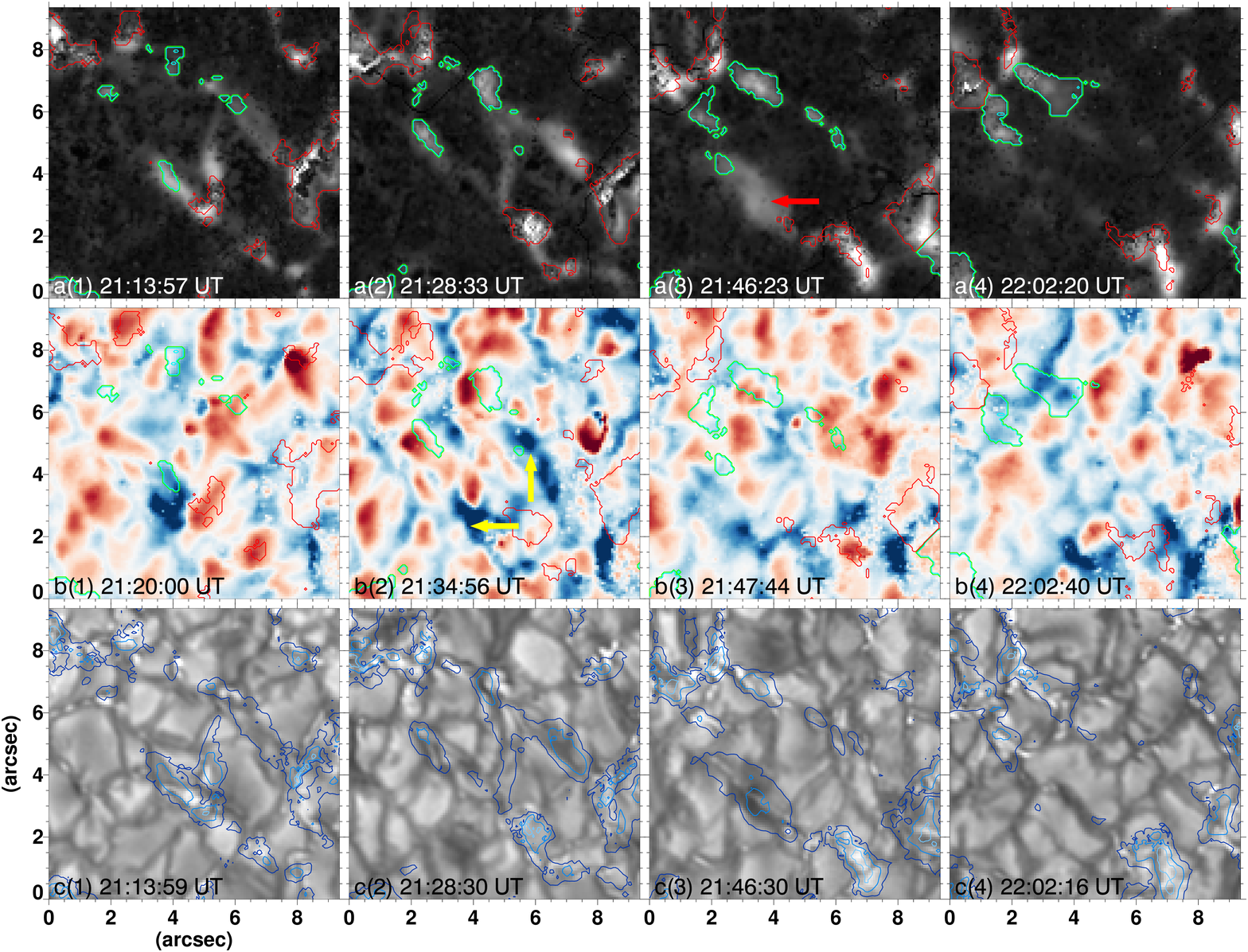}
    \caption{\textbf{Horizontal field and dopplergrams in event 2.} 
    Panels (a) show horizontal field map superimposed with vertical field contours at level of 150 G.  
    Panels (b) show upflows (downflows) of Dopplergrams in blue (red) color. The line-of-sight component the correspondent velocity is in range of $\pm$3.0~\kms. 
    Panels (c) present TiO images superimposed with horizontal field contours at levels of 200~G and 400~G, indicated by dark and light blue, respectively.
    The green (red) contours in (a) and (b) represent magnetic elements of negative (positive) polarity at level of 150 G.
    The red arrow in (a3) indicates horizontal component of the magnetic loop. The yellow arrows in (b2) indicate Doppler blue-shifts between the magnetic footpoints. An animation the horizontal field and dopplergrams in event 2 is available in the online Journal.
    The animated Figure, which includes the same panels (a)-(c) shown here, runs from 21:03 to 22:16~UT. \label{fig:f9}}
\end{figure}

    Starting from 21:13 UT, a new pair of magnetic elements appear at \sm[4$\farcs$5,3$\farcs$5] (Figure \ref{fig:f4}(a1)). The concentrated magnetic elements of opposite polarities continue to strengthen as they separate (as shown in Figures \ref{fig:f4}(a1--a3)). 
    It is noticeable from vector maps that the horizontal field enhances in place with the emerged magnetic elements and connects the diverging footpoints. 
    A loop-like magnetic field structure is observed between the footpoints FP2 and FP3 at 21:46~UT, and the width of the field loop reaches \sm1\arcsec\ as observed for its horizontal field component (Figure \ref{fig:f9}(a3)). 
    There is no obvious granular elongation observed to be associated with this horizontal field enhancement, while a deformed granule is accompanied by a transient magnetic enhancement between the footpoints FP2 and FP3 (see Figure \ref{fig:f4}(a3)). 
    The diffuse field can also be observed in Figures \ref{fig:f9}(a1--a4), which show the horizontal field map superimposed with vertical field contours at the level of 150~G. The green and red contours represent negative and positive magnetic elements, respectively.
    The Dopplergrams in Figures \ref{fig:f9}(b1--b4) show obvious red-shifts at footpoints and two blue-shifted patches connecting the footpoints at \sm21:35~UT. This indicates that the loop between footpoints has an upward motion and the footpoints have downward flows. The upflow speed reaches up to 1.8~\kms. 
    The emerging magnetic footpoints start to cancel with the preexisting magnetic fields of opposite polarities from 21:46~UT. 
    Such configuration of the emerged magnetic arc and the nearby preexisting footpoints in the north of the region may indicate the emergence of an undulating field in the emergence on the photosphere. Adjoining footpoints of opposite polarities in the emergent undulating field can easily organize a U-shaped or $\Omega$-shaped bald patch. According to previous studies \citep[e.g.,][]{2004ApJ...614.1099P,2017ApJ...836...63T}, the photospheric locations of bald patches of serpentine magnetic fields are very likely to be associated with EBs. 
    In the event 2, we witnessed bald patch associated EBs between the footpoints FP1 and FP2 (see Figure \ref{fig:f4} a4 and c4), where the brightening in H$\alpha$ wing occurs when the magnetic concentrations of opposite polarities approach each other. 
    The separation of emerged magnetic footpoints eventually reaches a maximum distance of 5 Mm at 22:02~UT.
    In Figure \ref{fig:f4}(c4), H$\alpha$ brightenings at +1.0~\AA\ off-band are observed at the magnetic footpoints ([2\arcsec,7\arcsec]) of the emerging flux at 22:02 UT, when the magnetic flux cancellation occurs. 
    At the same time, one can clearly observe a brightening in H$\alpha$ $-$0.4~\AA\ centered at \sm[4\arcsec,3\arcsec] (Figure \ref{fig:f4}(d4)) between the magnetic footpoints. 

\begin{figure}
    \epsscale{1.2}
    \plotone{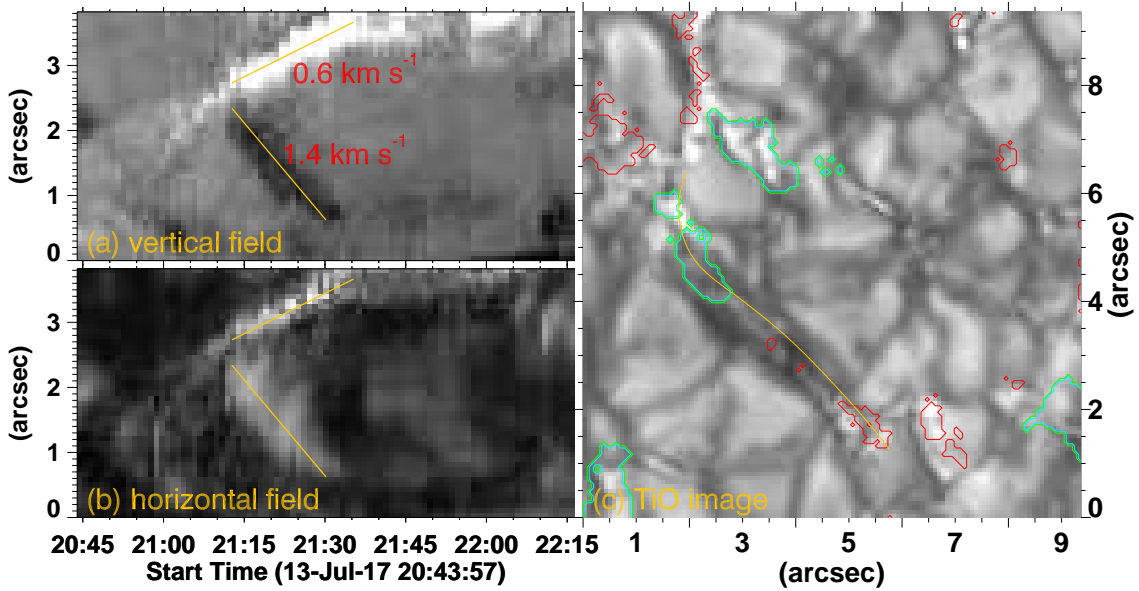}
    \caption{\textbf{Time-space diagrams of event 2.} 
    Panels (a) and (b) show time-space diagrams of vertical and horizontal field along the slit in the TiO image as shown in (c). Yellow lines in (a) and (b) trace and are used to estimate the speed of separation of the emerged magnetic polarities. Green (red) contours outline the magnetic elements of negative (positive) polarity. \label{fig:f5}}
\end{figure}

    The time-distance diagrams in Figure \ref{fig:f5} display bidirectional motions of the emerging magnetic elements. Similar phenomena were reported by \cite{Yang_2016} with TiO broadband filter images. The slit cuts along the extending magnetic loop as shown with the yellow curve in Figure \ref{fig:f5}(c). 
    Based on the time-distance diagrams, the magnetic footpoints diverge at a speed of 0.6--1.4~\kms, which is much slower than previous results \citep[3.8~\kms\ in][]{Yang_2016}. While the vertical fields follow confined separating traces, slightly weaker horizontal fields develop between the extending front of the horizontal field as seen in Figure \ref{fig:f5}(b). This is consistent with the observation from vector magnetic field maps that the magnetic footpoints are connected by diffused horizontal fields \citep{2007ApJ...666L.137C,2009ApJ...700.1391M}. 
    
\begin{figure}
    \epsscale{1.}
    \plotone{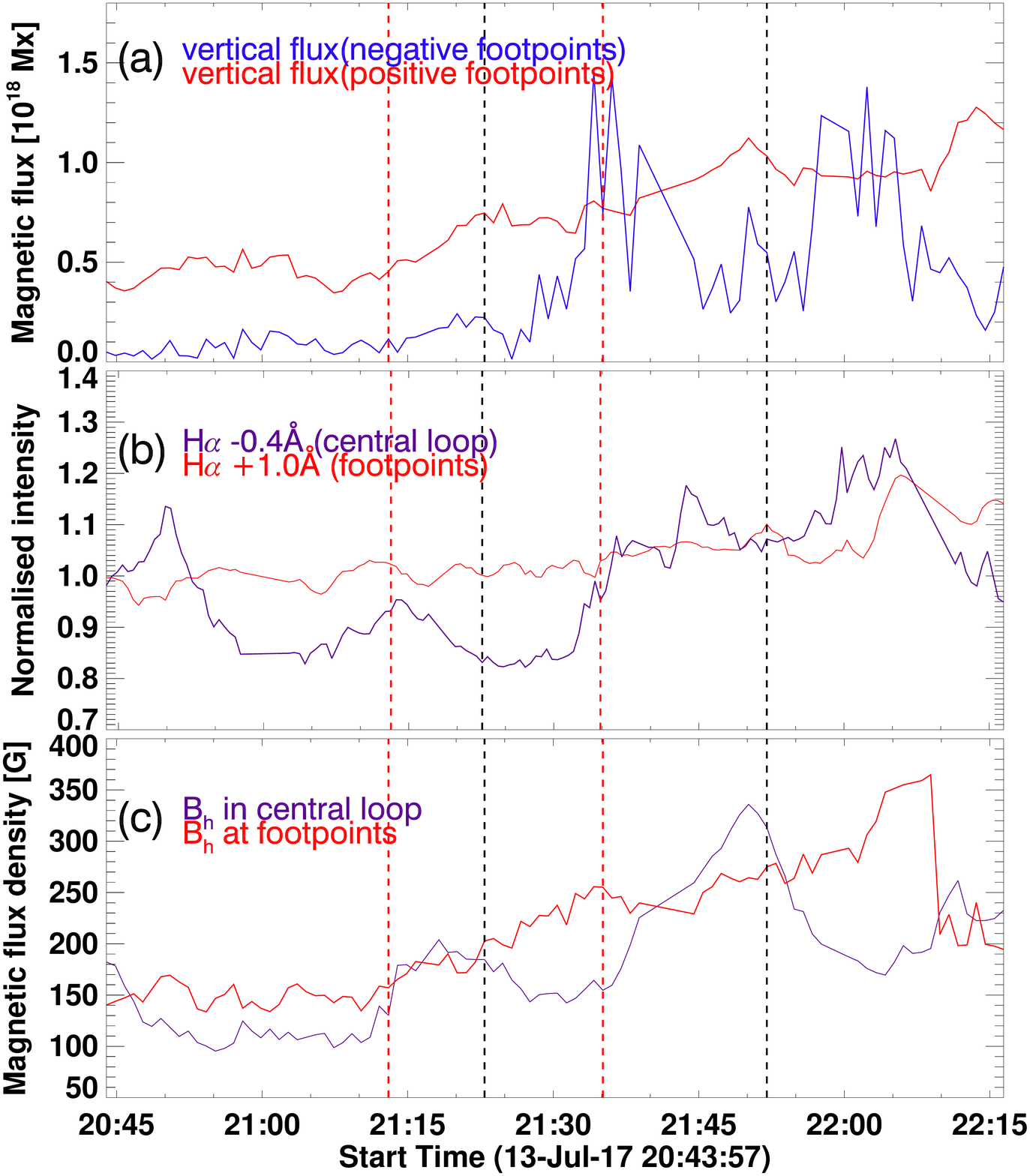}
    \caption{\textbf{The evolution of magnetic flux, mean brightness, and magnetic fields in event 2.} 
    Red and blue light curve in (a) shows averaged vertical flux evolution at footpoints FP3 and FP2 in Figure \ref{fig:f4}, respectively, in unit of $10^{18}$~Mx.
    Blue (red) light curve in (b) shows normalized intensity of H$\alpha$ $-$0.4 \AA (+1.0 \AA) in the loop (at footpoint FP1). 
    Blue (red) light curve in (c) shows horizontal field in the loop (footpoints) in unit of Gauss. Dashed lines in figure mark two episodes of flux emergence, 
    with red (black) dashed line represents start (end) time. 
\label{fig:f11}}
\end{figure}

    To understand the relationship between flux emergence and H$\alpha$ brightenings, we plot the temporal evolution of footpoint magnetic flux, H$\alpha$ intensities at $-$0.4~\AA\ and +1.0~\AA, and horizontal field strength in Figure \ref{fig:f11}. Figure \ref{fig:f11}(a) shows the averaged vertical flux in the positive (negative) footpoints as a red (blue) curve. 
    Figure \ref{fig:f11}(b) shows the normalized intensity of H$\alpha$ $-$0.4 and +1.0~\AA\ in the central loop (blue) and footpoint (red) regions, respectively.  
    Figure \ref{fig:f11}(c) shows averaged horizontal field in the same central loop (blue) and footpoint (red) regions.
    In the first phase of emergence, there is no visible H$\alpha$ response, while we observe brightenings in the loop corresponding to the second horizontal field increase starting from 21:46~UT.
    Comparing the light curves of the horizontal field at different locations, we find that the field strength increases at footpoints while decreases in the loop at \sm22:02~UT, which is co-temporal with H$\alpha$ brightenings. 
    Meanwhile, the vertical flux increases at the negative polarity footpoint. We speculate that H$\alpha$ brightenings in the loop are produced by the magnetic reconnection between the newly emerged magnetic loop with the overlying background field. On the other hand, the H$\alpha$ brightenings at footpoints are likely to be signatures of EBs between FP1 and FP2 (see Figure \ref{fig:f4}(a3)). 
    The LOS velocity maps of event 2 in Figure \ref{fig:f9}(b1)-(b4) show that the central loop and magnetic footpoints of the emerged flux loop is clearly associated with bi-directional shifts. At 21:13 UT, the velocity of blue-shift corresponding to the emerging loop is 0.45~\kms. It increases to 0.98~\kms\ at 21:28 UT then decreases to 0.37~\kms\ at 21:46 UT. The separating footpoints are observed to experience red-shifts with a maximum speed of 1.3~\kms\ at 21:28 UT.

\subsection{Properties of Other Events}\label{stat}
    Starting from 21:00 UT, with best-seeing quality of the day, we observe other small-scale flux emergence cases in \sm70~minutes, which demonstrate similar magnetic properties. 
    {The derived parameters of magnetic field evolution observed in nine events are given in Table \ref{tab:list}, including horizontal field, vertical field, vertical flux increments, the maximum distance of emerging bipolar magnetic elements, correspondent separation speed, LOS Doppler velocities, and associated EB occurrence. The maximum distance and correspondent average speed are measured in the emergence phase, which starts from the emergence of opposite polarities till both separation and flux enhancement cease.}
    As listed in Table \ref{tab:list}, five of the eight selected events in the observation can be categorized as a flux sheet type of emergence. We find that although the time interval between horizontal field emergence and the corresponding expanding granule boundaries is within 10~minutes, which is at the same time scale as summarized in previous studies \citep{2017ApJS..229....3C,2018ApJ...859L..26M}, magnetic elements in the granule boundaries continue to enhance as horizontal field increases and then either merge with adjacent magnetic fields or cancel with elements of opposite polarities. 
    The flux sheet emergence events 1 and 8 are observed to originate from intergranular dark lanes and form new expanding granular cells in the emergence locations. 
    While the other three emergent flux sheets (events 3, 7, and 9) do not show a direct linkage to pre-existing intergranular dark lanes, they are found to be located near the newly formed pores. 
    The vertical flux brought into the solar surface through emergence, which is associated with the expanding granules, is in the range of 0.9--11.6$\times$~$10^{18}$~Mx. As the edge of the emerging magnetic field that envelopes the granule expands at a speed of 1.5~\kms, the granule cells {undergoing emergence are averaged 4$\farcs$.3, which} grow by 0.7--1.5\arcsec. Although we observed a close connection between magnetic flux emergence and changes of photospheric granule structure, H$\alpha$ brightenings are rarely observed to be associated with flux sheet emergence. H$\alpha$ bright bursts captured in the event 9 region are closely associated with magnetic flux cancellation starting from 20:16 UT. During its emergence, TiO brightening at the granular boundary is observed at 21:36 UT.
    
    Summarizing the flux loop cases, we find that the vertical flux enhancement in this type of events is 3.0$\pm$0.9$\times$~$10^{18}$~Mx.while the separation speed of the emerging loop footpoints is 1.2~\kms, which is similar to the expanding speed of horizontal field in flux sheet emergence, {the maximum distance of opposite polarities reaches 5.5$\pm$1.5\arcsec. The difference of maximum separation is consistent with flux sheet and loop topology as magnetic footpoints of emerging flux loops are expected to extend further in the granular network.} Despite that H$\alpha$ brightenings are observed in event 2 at end of the flux emergence, the most prominent H$\alpha$ response occurred 38 minutes later. 
    In the other two flux loop emergence events (event 4 and 6), we also observed H$\alpha$ brightenings close to the emerged footpoints of these two events, while time intervals between emergence and H$\alpha$ brightenings do not show a similarity. In event 6 H$\alpha$ brightenings are observed three minutes after loop emergence. 
    {Among the studied events, five events are spatially associated with H$\alpha$ brightenings, including all three flux loop emergence and two flux sheet emergence.}
    From Doppler velocity maps of the flux emergence events, we find that the active region generally shows an upflow of 0.8~\kms\ in the background. Three of the listed emergence events (events 3-5) have blue-shifts over 2~\kms, which is comparable to previous observational results of photospheric Doppler velocity \citep{Ortiz_2014}. Event 5 is excluded from the categorization of magnetic topology because azimuthal ambiguity is not well resolved at the event location and Doppler red-shift is observed between opposite polarities. It is interpreted as a U-shaped field.

\section{summary and discussions}\label{sum}
    In this paper, we have presented a detailed study of small-scale flux emergence near the central PIL of NOAA AR 12665 on 2017 July 13. 
    The study is particularly focused on magnetic characteristics of two different kinds of flux emergence derived using the near-infrared polarimetric data obtained by NIRIS at BBSO/GST. 
    In addition, we studied photospheric evolution and chromospheric responses to the flux emergence using TiO and H$\alpha$ time-sequence images. 
    Our main results are summarized  below. 
    \begin{enumerate}
        \item In event 1, a typical sheet emergence case, an organized sheet-like structure of enhancing horizontal magnetic flux is seen to span over an entire granule, which expands at a speed of 1.6~\kms. The magnitude of the horizontal field in the flux sheet increases for \sm20~minutes, reaching up to 450 G. The emerged flux at footpoints reaches \sm1.8$\times$~$10^{18}$~Mx. In a subsequent second stage, the negative polarity footpoints and the co-spatial TiO bright points move along the intergranular lanes at a speed of \sm2~\kms.
        \item In event 2, a typical loop emergence case, magnetic footpoints at the two ends (the concentrated opposite-polarity flux component) emerge and move in the intergranular lanes with a separation speed of 1.2--1.7~\kms; meanwhile, a horizontal field lying in-between enhances, forming elongated, loop-like structures (the central diffused component). The positive vertical flux increases by \sm0.5$\times$~$10^{18}$~Mx. Later at \sm22:00~UT, horizontal field decreases in the central loop region while it increases at footpoint regions.
        \item Analysis of extended samples shows that all the eight events have a strongly emerged horizontal field of \sm450~G at maximum. 
        {While in the flux sheet emergence vertical field is comparable with the horizontal field(\sm270~G), in the loop emergence vertical field is 120~G stronger than the horizontal field.}
        In the five flux sheet emergence events, the horizontal field enhances and hovers the emergent granule cells as the granules grow. The concentration of field strength in the granule boundaries at the late phase of the emergence is observed in both horizontal and vertical magnetograms. Three out of the eight emergence events are observed to have a magnetic loop topology, in which the emergence of magnetic elements happens in intergranular lanes. The loop-like emergence carries \sm$10^{18}$~Mx of flux to the surface.   
    \end{enumerate}

    The results of the two types of flux emergence, with one experiencing an enhanced horizontal field hovering over the granule and the other following the typical $\Omega$--loop configuration, have advanced our understandings of small-scale flux emergence and formation of active regions. 
    It is worth noting that observations of flux-sheet emergence in both active regions \citep{2017ApJS..229....3C} and quiet Sun \citep{2019A&A...622L..12F} are rare. 
    The numerical study by \cite{2018ApJ...859L..26M} suggested that the occurrence rate of loop-like emergence (1--3~day$^{-1}$~Mm$^{-2}$) is \sm3 times higher than that of the sheet-like events (0.3--1~day$^{-1}$~Mm$^{-2}$) in the quiet Sun. In our study, we found more frequent occurrence of flux-sheet emergence events (1.8$\pm$0.1~day$^{-1}$~Mm$^{-2}$) than of loop-like emergence (1.1$\pm$0.06~day$^{-1}$~Mm$^{-2}$). We suspect that in the active region sub-surface magnetic tubes rising up to solar surface can break their original bipolar structure and emerge sideways due to the active and dynamic transverse motions. Frequent granulation observed in the active region provides higher opportunity than in quiet Sun to have magnetic tubes emerge with growing granules, which eventually form an emerging flux sheet.
    In comparison with a previous study, \cite{2019A&A...622L..12F} observed that the transverse flux density reaches up to 194~Mx$cm^{-2}$, corresponding to a maximum horizontal field of \sm300~G. 
    Our results show that the horizontal field reaches up to 450~G while the total flux is comparable to previous studies. Based on our results, five out of the eight observed flux emergence episodes in the FOV follow the flux-sheet type of emergence, and the rest follows the loop type emergence. Further, the flux sheets often appear in the emergence sites that are closely associated with newly evolving granulations. 
    Such a preference leads us to speculate that not only magnetic buoyancy instability but also transverse tension contribute to the flux-sheet emergence. 
    In both types of flux emergence, the {maximum distance of footpoint separation and} speed of Doppler shift vary with cases. Base on the results of our analyzed events, we conclude that despite differences in magnetic field topology and field strength distribution, the flux sheet and flux loop emergences share some similarities in terms of the emerging process. As an indication of Ellerman bombs, H$\alpha$ brightenings in our observations are found to have a close connection with magnetic loop emergence, in which the migrating footpoints collide and cancel with elements of opposite polarity in the intergranular lanes.
    
    The magnetic-loop emergences observed by us may evolve in the form of an undulating serpentine field. The three confirmed loop type emergences are observed in the magnetic intranetwork. 
    As magnetic footpoints diverge along the intergranular lanes, the emergent horizontal field is observed to enhance the field strength of network in magnetograms with correspondent dark lanes seen in TiO images. 
    Despite different emergence topology, the total emerged magnetic flux in the loop emergence events is comparable with that in the flux-sheet emergence events, and is an order of magnitude higher than previous studies of granule-sized magnetic loops \citep{2010A&A...511A..14G}. 
    As presented in the sample event 2, the magnetic footpoints of opposite polarities originate within neighboring granules and move apart along the intergranular lanes. 
    Thus as they approach the adjacent footpoints of the emerged field, a U-shaped field line can be formed across the surface. 
    Such magnetic field configuration is one type of bald patches that are found to have a strong connection with EBs \citep{2004ApJ...614.1099P,2017RAA....17...93J}.
    \cite{2015ApJ...812...11V} found that similar to EBs, flaring arch filaments could also exist in the emerging active region but are often observed as brightenings at H$\alpha$ core. 
    This phenomenon is believed to be related to the reconnection of curved fields.
    In comparison, our results in Section \ref{sub.2} reveal H$\alpha$ brightenings at the central loop location (in $-$0.4~\AA) as well as at the footpoints (in +1.0~\AA). 
    These may be interpreted as the reconnection between the emerging flux loop (footpoints) with the pre-existing overlying field (opposite polarity elements). 
    
    In summary, with high-resolution and high-cadence vector data, we have studied small-scale flux emergence from the observational perspective. We confirm that magnetic fields of granule-sized flux emergence have two different topologies, magnetic loop and flux sheet. 
    The primary difference of magnetic properties between the two types of emergence is that the magnetic field of flux sheets tend to be more inclined than arched magnetic loops.
    In association with the flux emergences, H$\alpha$ brightenings are more favorable to the footpoints of the emerging magnetic loops.
    Also, despite their different locations in the observed AR, both types of emergence bring 1--6$\times$~$10^{18}$~Mx of flux to the solar atmosphere.

    We thank the BBSO team for providing the data.
    BBSO operation is supported by NJIT and US NSF AGS-1821294 grant. GST operation is partly supported by the Korea Astronomy and Space Science Institute, the Seoul National University, and the Key Laboratory of Solar Activities of Chinese Academy of Sciences (CAS) and the Operation, Maintenance and Upgrading Fund of CAS for Astronomical Telescopes and Facility Instruments. 
    This work is supported by NSF under grants AGS-1821294, AGS-1954737, and AGS-1927578, by NASA under grants 80NSSC19K0257 and 80NSSC20K0025, and by National Science Foundation of China (NSFC) under grant 11729301.
    We also thank the referee for the constructive comments and suggestions that greatly help improve this paper.

\facilities{Big Bear Solar Observatory}
\bibliography{reference}

\begin{thebibliography}{}
\expandafter\ifx\csname natexlab\endcsname\relax\def\natexlab#1{#1}\fi
\providecommand{\url}[1]{\href{#1}{#1}}
\providecommand{\dodoi}[1]{doi:~\href{http://doi.org/#1}{\nolinkurl{#1}}}
\providecommand{\doeprint}[1]{\href{http://ascl.net/#1}{\nolinkurl{http://ascl.net/#1}}}
\providecommand{\doarXiv}[1]{\href{https://arxiv.org/abs/#1}{\nolinkurl{https://arxiv.org/abs/#1}}}

\bibitem[{{Bello Gonz{\'a}lez} {et~al.}(2013){Bello Gonz{\'a}lez}, {Danilovic},
  \& {Kneer}}]{2013A&A...557A.102B}
{Bello Gonz{\'a}lez}, N., {Danilovic}, S., \& {Kneer}, F. 2013, \aap, 557,
  A102, \dodoi{10.1051/0004-6361/201321632}

\bibitem[{{Campos Rozo} {et~al.}(2019){Campos Rozo}, {Utz}, {Vargas
  Dom{\'\i}nguez}, {Veronig}, \& {Van Doorsselaere}}]{2019A&A...622A.168C}
{Campos Rozo}, J.~I., {Utz}, D., {Vargas Dom{\'\i}nguez}, S., {Veronig}, A., \&
  {Van Doorsselaere}, T. 2019, \aap, 622, A168,
  \dodoi{10.1051/0004-6361/201832760}

\bibitem[{{Cao} {et~al.}(2012){Cao}, {Goode}, {Ahn}, {Gorceix}, {Schmidt}, \&
  {Lin}}]{2012ASPC..463..291C}
{Cao}, W., {Goode}, P.~R., {Ahn}, K., {et~al.} 2012, in Astronomical Society of
  the Pacific Conference Series, Vol. 463, Second ATST-EAST Meeting: Magnetic
  Fields from the Photosphere to the Corona., 291

\bibitem[{{Centeno}(2012)}]{2012ApJ...759...72C}
{Centeno}, R. 2012, \apj, 759, 72, \dodoi{10.1088/0004-637X/759/1/72}

\bibitem[{{Centeno} {et~al.}(2007){Centeno}, {Socas-Navarro}, {Lites}, {Kubo},
  {Frank}, {Shine}, {Tarbell}, {Title}, {Ichimoto}, {Tsuneta}, {Katsukawa},
  {Suematsu}, {Shimizu}, \& {Nagata}}]{2007ApJ...666L.137C}
{Centeno}, R., {Socas-Navarro}, H., {Lites}, B., {et~al.} 2007, \apjl, 666,
  L137, \dodoi{10.1086/521726}

\bibitem[{{Centeno} {et~al.}(2017){Centeno}, {Blanco Rodr{\'\i}guez}, {Del Toro
  Iniesta}, {Solanki}, {Barthol}, {Gand orfer}, {Gizon}, {Hirzberger},
  {Riethm{\"u}ller}, {van Noort}, {Orozco Su{\'a}rez}, {Berkefeld}, {Schmidt},
  {Mart{\'\i}nez Pillet}, \& {Kn{\"o}lker}}]{2017ApJS..229....3C}
{Centeno}, R., {Blanco Rodr{\'\i}guez}, J., {Del Toro Iniesta}, J.~C., {et~al.}
  2017, \apjs, 229, 3, \dodoi{10.3847/1538-4365/229/1/3}

\bibitem[{{Cheung} {et~al.}(2007){Cheung}, {Sch{\"u}ssler}, \&
  {Moreno-Insertis}}]{2007A&A...467..703C}
{Cheung}, M.~C.~M., {Sch{\"u}ssler}, M., \& {Moreno-Insertis}, F. 2007, \aap,
  467, 703, \dodoi{10.1051/0004-6361:20077048}

\bibitem[{{De Pontieu}(2002)}]{2002ApJ...569..474D}
{De Pontieu}, B. 2002, \apj, 569, 474, \dodoi{10.1086/339231}

\bibitem[{{DeForest} {et~al.}(2007){DeForest}, {Hagenaar}, {Lamb}, {Parnell},
  \& {Welsch}}]{2007ApJ...666..576D}
{DeForest}, C.~E., {Hagenaar}, H.~J., {Lamb}, D.~A., {Parnell}, C.~E., \&
  {Welsch}, B.~T. 2007, \apj, 666, 576, \dodoi{10.1086/518994}

\bibitem[{{Ellerman}(1917)}]{1917ApJ....46..298E}
{Ellerman}, F. 1917, \apj, 46, 298, \dodoi{10.1086/142366}

\bibitem[{{Fan}(2001)}]{2001ApJ...554L.111F}
{Fan}, Y. 2001, \apjl, 554, L111, \dodoi{10.1086/320935}

\bibitem[{{Fischer} {et~al.}(2019){Fischer}, {Borrero}, {Bello Gonz{\'a}lez},
  \& {Kaithakkal}}]{2019A&A...622L..12F}
{Fischer}, C.~E., {Borrero}, J.~M., {Bello Gonz{\'a}lez}, N., \& {Kaithakkal},
  A.~J. 2019, \aap, 622, L12, \dodoi{10.1051/0004-6361/201834628}

\bibitem[{{G{\"o}m{\"o}ry} {et~al.}(2010){G{\"o}m{\"o}ry}, {Beck}, {Balthasar},
  {Ryb{\'a}k}, {Ku{\v{c}}era}, {Koza}, \& {W{\"o}hl}}]{2010A&A...511A..14G}
{G{\"o}m{\"o}ry}, P., {Beck}, C., {Balthasar}, H., {et~al.} 2010, \aap, 511,
  A14, \dodoi{10.1051/0004-6361/200912807}

\bibitem[{{Goode} \& {Cao}(2012)}]{2012SPIE.8444E..03G}
{Goode}, P.~R., \& {Cao}, W. 2012, in \procspie, Vol. 8444, Ground-based and
  Airborne Telescopes IV, 844403, \dodoi{10.1117/12.925494}

\bibitem[{{Jiang} {et~al.}(2017){Jiang}, {Feng}, {Wu}, \&
  {Hu}}]{2017RAA....17...93J}
{Jiang}, C.-W., {Feng}, X.-S., {Wu}, S.-T., \& {Hu}, Q. 2017, Research in
  Astronomy and Astrophysics, 17, 093, \dodoi{10.1088/1674-4527/17/9/93}

\bibitem[{Jin {et~al.}(2008)Jin, Wang, \& Zhao}]{Jin_2008}
Jin, C., Wang, J., \& Zhao, M. 2008, The Astrophysical Journal, 690, 279,
  \dodoi{10.1088/0004-637x/690/1/279}

\bibitem[{{Leka} {et~al.}(2009{\natexlab{a}}){Leka}, {Barnes}, \&
  {Crouch}}]{2009ASPC..415..365L}
{Leka}, K.~D., {Barnes}, G., \& {Crouch}, A. 2009{\natexlab{a}}, in
  Astronomical Society of the Pacific Conference Series, Vol. 415, The Second
  Hinode Science Meeting: Beyond Discovery-Toward Understanding, ed.
  B.~{Lites}, M.~{Cheung}, T.~{Magara}, J.~{Mariska}, \& K.~{Reeves}, 365

\bibitem[{{Leka} {et~al.}(2009{\natexlab{b}}){Leka}, {Barnes}, {Crouch},
  {Metcalf}, {Gary}, {Jing}, \& {Liu}}]{2009SoPh..260...83L}
{Leka}, K.~D., {Barnes}, G., {Crouch}, A.~D., {et~al.} 2009{\natexlab{b}},
  \solphys, 260, 83, \dodoi{10.1007/s11207-009-9440-8}

\bibitem[{{Lim} {et~al.}(2011){Lim}, {Yurchyshyn}, {Abramenko}, {Ahn}, {Cao},
  \& {Goode}}]{2011ApJ...740...82L}
{Lim}, E.-K., {Yurchyshyn}, V., {Abramenko}, V., {et~al.} 2011, \apj, 740, 82,
  \dodoi{10.1088/0004-637X/740/2/82}

\bibitem[{{Lites} {et~al.}(1996){Lites}, {Leka}, {Skumanich}, {Martinez
  Pillet}, \& {Shimizu}}]{1996ApJ...460.1019L}
{Lites}, B.~W., {Leka}, K.~D., {Skumanich}, A., {Martinez Pillet}, V., \&
  {Shimizu}, T. 1996, \apj, 460, 1019, \dodoi{10.1086/177028}

\bibitem[{{Mart{\'\i}nez Gonz{\'a}lez} \& {Bellot
  Rubio}(2009)}]{2009ApJ...700.1391M}
{Mart{\'\i}nez Gonz{\'a}lez}, M.~J., \& {Bellot Rubio}, L.~R. 2009, \apj, 700,
  1391, \dodoi{10.1088/0004-637X/700/2/1391}

\bibitem[{{Metcalf}(1994)}]{1994SoPh..155..235M}
{Metcalf}, T.~R. 1994, \solphys, 155, 235, \dodoi{10.1007/BF00680593}

\bibitem[{{Moreno-Insertis} {et~al.}(2018){Moreno-Insertis}, {Martinez-Sykora},
  {Hansteen}, \& {Mu{\~n}oz}}]{2018ApJ...859L..26M}
{Moreno-Insertis}, F., {Martinez-Sykora}, J., {Hansteen}, V.~H., \&
  {Mu{\~n}oz}, D. 2018, \apjl, 859, L26, \dodoi{10.3847/2041-8213/aac648}

\bibitem[{{Oba} {et~al.}(2017){Oba}, {Riethm{\"u}ller}, {Solanki}, {Iida},
  {Quintero Noda}, \& {Shimizu}}]{2017ApJ...849....7O}
{Oba}, T., {Riethm{\"u}ller}, T.~L., {Solanki}, S.~K., {et~al.} 2017, \apj,
  849, 7, \dodoi{10.3847/1538-4357/aa8e44}

\bibitem[{Ortiz {et~al.}(2014)Ortiz, Rubio, Hansteen, de~la
  Cruz~Rodr{\'{\i}}guez, \& van~der Voort}]{Ortiz_2014}
Ortiz, A., Rubio, L. R.~B., Hansteen, V.~H., de~la Cruz~Rodr{\'{\i}}guez, J.,
  \& van~der Voort, L.~R. 2014, The Astrophysical Journal, 781, 126,
  \dodoi{10.1088/0004-637x/781/2/126}

\bibitem[{{Otsuji} {et~al.}(2011){Otsuji}, {Kitai}, {Ichimoto}, \&
  {Shibata}}]{2011PASJ...63.1047O}
{Otsuji}, K., {Kitai}, R., {Ichimoto}, K., \& {Shibata}, K. 2011, \pasj, 63,
  1047, \dodoi{10.1093/pasj/63.5.1047}

\bibitem[{{Pariat} {et~al.}(2004){Pariat}, {Aulanier}, {Schmieder},
  {Georgoulis}, {Rust}, \& {Bernasconi}}]{2004ApJ...614.1099P}
{Pariat}, E., {Aulanier}, G., {Schmieder}, B., {et~al.} 2004, \apj, 614, 1099,
  \dodoi{10.1086/423891}

\bibitem[{{Parker}(1977)}]{1977ARA&A..15...45P}
{Parker}, E.~N. 1977, \araa, 15, 45,
  \dodoi{10.1146/annurev.aa.15.090177.000401}

\bibitem[{{Schmidt}(1968)}]{1968IAUS...35...95S}
{Schmidt}, H.~U. 1968, in IAU Symposium, Vol.~35, Structure and Development of
  Solar Active Regions, ed. K.~O. {Kiepenheuer}, 95

\bibitem[{{Schmidt}(1974)}]{1974IAUS...56...35S}
{Schmidt}, H.~U. 1974, in IAU Symposium, Vol.~56, Chromospheric Fine Structure,
  ed. R.~G. {Athay}, 35

\bibitem[{{Shumko} {et~al.}(2014){Shumko}, {Gorceix}, {Choi}, {Kellerer},
  {Cao}, {Goode}, {Abramenko}, {Richards}, {Rimmele}, \&
  {Marino}}]{2014SPIE.9148E..35S}
{Shumko}, S., {Gorceix}, N., {Choi}, S., {et~al.} 2014, in \procspie, Vol.
  9148, Adaptive Optics Systems IV, 914835, \dodoi{10.1117/12.2056731}

\bibitem[{{Smitha} {et~al.}(2017){Smitha}, {Anusha}, {Solanki}, \&
  {Riethm{\"u}ller}}]{2017ApJS..229...17S}
{Smitha}, H.~N., {Anusha}, L.~S., {Solanki}, S.~K., \& {Riethm{\"u}ller}, T.~L.
  2017, \apjs, 229, 17, \dodoi{10.3847/1538-4365/229/1/17}

\bibitem[{{Tian} {et~al.}(2010){Tian}, {Yao}, {Zong}, {He}, \&
  {Qi}}]{2010ApJ...720..454T}
{Tian}, H., {Yao}, S., {Zong}, Q., {He}, J., \& {Qi}, Y. 2010, \apj, 720, 454,
  \dodoi{10.1088/0004-637X/720/1/454}

\bibitem[{{Toriumi} {et~al.}(2017){Toriumi}, {Katsukawa}, \&
  {Cheung}}]{2017ApJ...836...63T}
{Toriumi}, S., {Katsukawa}, Y., \& {Cheung}, M.~C.~M. 2017, \apj, 836, 63,
  \dodoi{10.3847/1538-4357/836/1/63}

\bibitem[{{Vissers} {et~al.}(2015){Vissers}, {Rouppe van der Voort}, {Rutten},
  {Carlsson}, \& {De Pontieu}}]{2015ApJ...812...11V}
{Vissers}, G.~J.~M., {Rouppe van der Voort}, L.~H.~M., {Rutten}, R.~J.,
  {Carlsson}, M., \& {De Pontieu}, B. 2015, \apj, 812, 11,
  \dodoi{10.1088/0004-637X/812/1/11}

\bibitem[{Vissers {et~al.}(2013)Vissers, van~der Voort, \&
  Rutten}]{Vissers_2013}
Vissers, G. J.~M., van~der Voort, L. H. M.~R., \& Rutten, R.~J. 2013, The
  Astrophysical Journal, 774, 32, \dodoi{10.1088/0004-637x/774/1/32}

\bibitem[{{Wang} {et~al.}(2017){Wang}, {Liu}, {Ahn}, {Xu}, {Jing}, {Deng},
  {Huang}, {Liu}, {Kusano}, {Fleishman}, {Gary}, \&
  {Cao}}]{2017NatAs...1E..85W}
{Wang}, H., {Liu}, C., {Ahn}, K., {et~al.} 2017, Nature Astronomy, 1, 0085,
  \dodoi{10.1038/s41550-017-0085}

\bibitem[{Watanabe {et~al.}(2011)Watanabe, Vissers, Kitai, van~der Voort, \&
  Rutten}]{Watanabe_2011}
Watanabe, H., Vissers, G., Kitai, R., van~der Voort, L.~R., \& Rutten, R.~J.
  2011, The Astrophysical Journal, 736, 71, \dodoi{10.1088/0004-637x/736/1/71}

\bibitem[{{W{\"o}ger} {et~al.}(2008){W{\"o}ger}, {von der L{\"u}he}, \&
  {Reardon}}]{2008A&A...488..375W}
{W{\"o}ger}, F., {von der L{\"u}he}, O., \& {Reardon}, K. 2008, \aap, 488, 375,
  \dodoi{10.1051/0004-6361:200809894}

\bibitem[{Yang {et~al.}(2016)Yang, Chae, Lim, Song, Cho, Kwak, Yurchyshyn, \&
  Kim}]{Yang_2016}
Yang, H., Chae, J., Lim, E.-K., {et~al.} 2016, The Astrophysical Journal, 829,
  100, \dodoi{10.3847/0004-637x/829/2/100}

\bibitem[{{Yurchyshyn} {et~al.}(2012){Yurchyshyn}, {Ahn}, {Abramenko}, {Goode},
  \& {Cao}}]{2012arXiv1207.6418Y}
{Yurchyshyn}, V., {Ahn}, K., {Abramenko}, V., {Goode}, P., \& {Cao}, W. 2012,
  arXiv e-prints, arXiv:1207.6418.
\newblock \doarXiv{1207.6418}

\bibitem[{{Zwaan}(1985)}]{1985SoPh..100..397Z}
{Zwaan}, C. 1985, \solphys, 100, 397, \dodoi{10.1007/BF00158438}

\end{thebibliography}
\bibliographystyle{aasjournal}

\begin{deluxetable*}{cccccccc}
    \tablenum{1}
    \tablecaption{Magnetic Properties of the Observed Events\label{tab:list}}
    \tablewidth{0pt}
    \tablehead{
    \colhead{Event} & \colhead{Horizontal} & \colhead{Vertical} &
    \colhead{Flux} & \colhead{Maximum} & \colhead{Separation} & \colhead{Doppler} & \colhead{EBs occurrence} \\
    \colhead{Number} & \colhead{Field (G)} & \colhead{Field (G)} &
    \colhead{($\times$~$10^{18}$~Mx)} & \colhead{Distance (\arcsec)} & \colhead{Speed (\kms)} & \colhead{V (\kms)} & \colhead{(Y/N)}
    }
    \decimalcolnumbers
    \startdata
    1$\star$ & 390/280 & 250/250 & 4.4/2.0 & 4.4 & 1.6 & 0.45 & N \\
    2$\diamond$ & 180/148 & 500/298 & 1.0/1.2 & 7 & 1.3 & 0.98 & Y \\
    3$\star$ & 378/225 & 320 & 1.7/2.5 & 3.3 & 0.9 & 2.53 & N \\
    4$\diamond$ & 280/200 & 435/150 & 1.9/0.39 & 4.2 & 1.5 & 2.45 & Y \\
    5 & 360/230 & 400 & 3.8/0.6 & 3 & 2.0 & 2.64 & Y \\
    6$\diamond$ & 328/240 & 530 & 5.8/6 & 5.5 & 1.2 & 1.70 & Y \\
    7$\star$ & 303/155 & 220$\pm$40 & 1.29/0.98 & 4.3 & 1.8 & 1.47 & N \\
    8$\star$ & 425/318 & 310/574 & 8.6/5.6 & 6 & 3.5 & 0.9 & N \\
    9$\star$ & 500/350 & 260 & 0.98 & 3.8 & / & 0.64 & N \\
    \enddata
    \tablecomments{Flux sheet emergence events are labeled as ($\star$) and flux loop emergence events are labeled as ($\diamond$) after event numbers. Maximum/average field strengths of each event are presented in columns (2) and (3). Positive/negative vertical flux increments of through the emergence are presented in column (4). Maximum distances and speed of oppsite polarity separation in the emergence phase are presented in column (5) and (6), respectively. LOS Doppler upflow speeds are presented in column (7). Emergence event associated with EB observation in H$\alpha$ is labeled as Y and emergence without EB association is labeled as N in column (8). Event 5 is excluded from discussion in Section \ref{stat}.} 
\end{deluxetable*}

    \end{document}